\documentclass[11pt,twoside]{article}   
\usepackage{epsfig}
\begin{document}           
\begin{center}{\bf MECO In An Exponential Metric}

\bigskip

Stanley L. Robertson \footnote{Dept. of Chemistry \& Physics, Emeritus, Southwestern Oklahoma State University\\
Weatherford, OK 73096,
 (stanrobertson@itlnet.net)}

\begin{abstract}
Magnetic Eternally Collapsing Objects (MECO) have been proposed as
the central engines of galactic black hole candidates (GBHC) and
supermassive active galactic nuclei (AGN). Previous work has shown
that their luminosities and spectral and timing characteristics
are in good agreement with observations. These features and the
formation of jets are generated primarily by the interactions of
accretion disks with an intrinsically magnetic central MECO. The
interaction of accretion disks with the anchored magnetic fields
of the central objects permits a unified description of properties
for GBHC, AGN, neutron stars in low mass x-ray binaries and dwarf
novae systems. The previously published MECO models have been
based on a quasistatic Schwarzschild metric of General Relativity;
however, the only essential feature of this metric is its ability
to produce extreme gravitational redshifts. For reasons discussed
in this article, an alternative development based on a quasistatic
exponential metric is considered here.

\end{abstract}
\end{center}

\section{Introduction}
Stellar mass objects compact enough to be black holes are
ubiquitous in galaxies while supermassive compact objects are
found in the nuclei of most, if not all, galaxies. Although these
objects are routinely called black holes, no compelling evidence
of an event horizon, the quintessential feature of a black hole,
has yet been found. Though not widely considered, it is possible
that these objects may not possess event horizons at all. They
might instead be some kind of eternally collapsing object (ECO)
that exhibits extreme gravitational redshift and consequently,
very low luminosity.

Once an object becomes smaller than its photon sphere, outgoing
radiation begins to be trapped; the more compact it becomes, the
more efficient the trap. Within the scope of general relativity,
it has been shown (Herrera \& Santos 2004, Herrera, DiPrisco \&
Barreto 2006, Mitra 2006a,b,c, Mitra \& Glendenning 2010,
Robertson \& Leiter 2004) that it is possible to achieve an
Eddington balance before the formation of trapped surfaces or
event horizons. There is evidence that such non-black hole objects
exist and that they are strongly magnetic (Robertson \& Leiter
2002, 2003, 2004 hereafter RL02, RL03, RL04, Schild, Leiter \&
Robertson 2006, 2008, Schild \& Leiter 2010) The Robertson-Leiter
MECO variant adds a (M)agnetic field that can reach a quantum
electrodynamic limit. It contributes heavily to the outgoing
photon flow (Robertson \& Leiter 2006, 2010 hereafter RL06, RL10).
This eventually allows all of the observable properties of their
MECO model to depend only upon two parameters - mass and spin, but
MECO models with other magnetic field configurations might also be
of interest. Despite the high surface luminosity of an Eddington
balance, distantly observed MECO are very low luminosity objects
capable of satisfying all observational constraints due to extreme
gravitational redshift of the photons that do escape.

There are strong observational signatures of the magnetic central
objects within Low Mass X-ray Binary systems, AGN and dwarf novae.
Low and high luminosity states, spectral state switches between
them, quasi-periodic oscillations and jet formation are common
features. These states and oscillations have been described in
terms of the interaction of accretion disks threaded by poloidal
magnetic fields (Chou\& Tajima 1999). Some very revealing
simulations of some of the process can be seen here\\
 {\bf http://astrosun2.astro.cornell.edu/us-rus/jets.htm }
These show clearly the formation of jets and funnel flows that can
be compared with observational data and correlations of
quasi-periodic oscillations of dwarf novae, neutron stars and GBHC
(Mauche 2002, Warner \& Woudt 2003)

In previous work (RL02, RL04, RL06) evidence was provided for the
existence of intrinsic magnetic moments of $\sim 10^{29-30}~ Gauss
~cm^3$ in the GBHC of LMXB. Others have reported evidence for
strong magnetic fields in GBHC. A field in excess of $10^8$ G has
been found at the base of the jets of GRS 1915+105 (Gliozzi, Bodo
\& Ghisellini 1999, Vadawale, Rao \& Chakrabarti 2001). Based on a
study of optical polarization of Cygnus X-1 in its low state
(Gnedin et al. 2003) has found a slow GBHC spin and a magnetic
field of $\sim 10^8 Gauss$ at the location of its optical
emission. Since these field strengths exceed disk plasma
equipartition levels they can be attributed to a magnetic central
object. They are in good agrement with the magnetic moments that
have previously been presented (RL06) It was also shown (RL04)
that the MECO model and its jet mechanism scales up to the AGN
without difficulty. The empty inner disk region shown by
microlensing of quasars (Schild, Leiter \& Robertson 2006, 2008,
Schild \& Leiter 2010) has been shown to be consistent with the
magnetic field structure of the MECO model. Lastly, it was shown
that the MECO model can reconcile the low luminosity of Sgr A*
with its expected Bondi accretion rate and account for the
orthogonal polarizations of near infrared and millimeter radio
emissions of Sgr A* (RL10).

The most attractive feature of the MECO model is that a central
object with an intrinsic magnetic field interacts with accretion
disks in a way that permits a unified description of the similar
observable characteristics of dwarf novae, LMXB neutron stars,
GBHC and AGN. While there are some observable differences between
neutron stars, dwarf novae and GBHC, it has not been demonstrated
that any of the differences can be attributed to the presence of
an event horizon. The only thing that is needed is a large
gravitational redshift to satisfy the requirements of low
luminosity. The success of the MECO models demonstrates clearly
that there is as yet no need of event horizons in astrophysics.
That being the case, it is of some interest to consider
alternatives that do not possess them. The essential ingredients
and details of MECO are recapitulated here for an exponential
metric that is capable of producing the required redshifts.

But with a successful model of MECO based on a Schwarzschild
metric one might ask why a different basis for MECO might be of
interest. The answer is that we hope to eventually settle the
question of whether or not black holes exist by comparing
observations and models with and without event horizons. A model
based on a metric with no event horizon might help to clarify the
issues. Stated another way, before the existence of black holes is
accepted it ought to be necessary to find black hole candidates
exhibiting characteristics that cannot be easily accommodated by a
metric that lacks an event horizon.

\section{The Exponential Metric}

Event horizons are problematic in their own right. The question of
how quantum mechanics and general relativity might be reconciled
has recently been sharpened by considering what happens to a
freely falling particle of matter approaching an event horizon.
The possibility that it might meet a radiative ``firewall" has
recently become a very active research topic (e.g., Abramowicz,
Kluzniak \& Lasota 2013, Anastopoulos \& Savvidou, 2014, Hawking
2014). Hawking says that event horizons cannot exist as stable
features of compact objects.

This is a problem of such importance that we should consider all
aspects; however, the necessity of event horizons in astrophysics
seems not to have been questioned. They first arose in the
solutions of the Einstein field equations applicable to the
exterior of a central compact mass and are manifest by the
vanishing of the time component of the metric tensor, which for
the Schwarzschild metric, is
\begin{equation}
g_{00}=1-2R_g/r~,~~~~~~~R_g=GM/c^2
\end{equation}

The gravitational redshift $z$, of photons leaving the surface of
a Schwarzschild metric compact object is given by $1+z =
g_{00}^{-1/2}$, which is infinite with $g_{00}=0$ at the
Schwarzschild radius, $R_s = 2R_g$. This occurrence of an event
horizon can be traced to Einstein's exclusion of gravitational
field energy density as a source term in his field equations. All
forms of mass-energy EXCEPT that of gravitational fields were
considered as sources. The (dimensionless) Newtonian gravitational
potential function $u(r) = GM/(c^2/r)$ is actually a solution of
the Einstein field equations if its gravitational field energy
given by $(c^2 \nabla u(r))^2/(8 \pi G)$ is used as a source term
in the right member of the Einstein field in the space exterior to
mass M (see Appendix B). In this case the event horizon disappears
as $g_{00}$ becomes $e^{-2u(r)}$, with a classical point particle
singularity at $r=0$ that is of no consequence. The field energy
density is analogous to the field energy density of a static
electromagnetic field. It is of second degree in the derivative of
the potential and is of second order in its affect on curvature.
If it is truly needed for a correct description of gravity, then
General Relativity would seem to be just a first order theory;
good enough for weak fields but not to be trusted in the strong
fields of objects as compact as $2R_g$.

If the gravitational field is a physical entity in its own right,
rather than being manifest solely by a spacetime curvature that
has excluded the effect of gravitational field energy, then one
might consider gravitational redshift to be a function of the
(dimensionless) gravitational potential $u(r)=GM/(c^2r)$.

Consider the frequencies of photons leaving the surface of a
compact mass, $M$, and being observed at locations a,b and c in
the external space. If the corresponding observed frequencies
would be $\nu_a, \nu_b, \nu_c$, then their ratios would be
functions of the potentials such as $\nu_a/\nu_b =
f(u(r_a),u(r_b))$. But since potentials are only determined to
within an additive constant, this would be the same as
$f(u(r_a)+C,u(r_b)+C)$.  So consider $\nu_a/\nu_b
=(\nu_a/\nu_c)(\nu_c/\nu_b)= f(u(r_a)+C,u(r_b)+C) =
f(u(r_a),u(r_c)) f(u(r_c),u(r_b))$.  If we would choose $C
=-u(r_b)$ then the first ratio would become $f(u(r_a)-u(r_b), 0)$.
Similarly we could add a constant $-u(r_c)$ to each of the next
two terms and obtain $f(u(r_a)-u(r_b), 0)= f(u(r_a)-,u(r_c))
f(0,u(r_b)-u(r_c))$. If we let $A=u(r_a)-u(r_c)$ and
$B=u(r_c)-u(r_b)$ we have $f(A+B),0)=f(A,0)f(0,-B)$. But since it
must also be possible to offset the potentials by a constant
amount $+B$, it follows that $f(0,-B)=f(B,0)$. So we arrive at
$f(A+B,0) = f(A,0)f(B,0)$. Thus we conclude that the observed
frequencies of photons should be proportional to an exponential
function, $f(u(r))=e^{u(r)}$, and the gravitational redshift
observed very distantly from a mass of radius $R$, should be given
by
\begin{equation}
1+z= e^{u(R)} = e^{R_g/R}
\end{equation}
It is of considerable interest to note that this result also
follows exactly from Einstein's principle of equivalence and the
special theory of relativity (see Appendix A).

 If we consider a meter stick transported from the compact object
surface out to various distances to be dependent on the
gravitational potential, we would conclude that its length must
also be an exponential function of the potential. If the
orientation of the stick relative to the gravitational field
direction would not matter, then we can describe the effects of
gravity in terms of the isotropic metric form:
\begin{equation}
ds^2 = e^{-2u(r)}c^2 dt^2 - e^{2u(r)}(dr^2+r^2d\theta^2+r^2
sin^2(\theta)d\phi^2)
\end{equation}
This is the generic metric that will be used for the remainder of
this article. It was first proposed by Yilmaz (1958, 1971), but
introduced with the preceding arguments by Rastall (1975). While
this metric is far, far from a complete gravity theory it can
encompass essentially all of relativistic astrophysics except
gravitational radiation and waves and large scale cosmology. Its
main appeal in the present case is that it can yield large and
arguably correct gravitational redshifts without an event horizon.

Several features of immediate interest pertain to geodesic motions
of particles in this exponential metric: (i) There is an innermost
marginally stable orbit at a distance of $5.24 R_g$ from the
central object center, which is only slightly smaller than the
$6R_g$ of the Schwarzschild metric. This small difference is due
to the isotropic form of the metric. (ii) The energy that can be
liberated in an accretion disk that reaches the marginally stable
orbit is 5.5\% of the rest mass energy of the particles, compared
to 5.7\% in the Schwarzschild metric. (iii) There is a circular
photon orbit and ``photon sphere" at $r=2R_g$, compared to $3R_g$
in the Schwarzschild metric. There are no stable closed orbits
that exist within or pass inside the photon sphere for particles
with nonzero rest mass, however, particles with sufficient energy
might escape from inside the photon sphere. (iv) Unlike a
Schwarzschild metric, there is no event horizon from which photons
cannot escape. Photons with purely radial motions can always
escape, however those not directed radially are increasingly
restricted to small escape cones.

Photons that would pass too near an object that would be small
enough to reside within its photon sphere would be captured. In
order to get past the central object, their path must come no
closer than the ``capture radius" from the center. This capture
radius is $\sqrt{27} R_g$ for the Schwarzschild metric and $2 e
R_g$ for the exponential metric. Since these differ by only 4.6\%,
the shadows cast by compact astronomical objects will probably not
be capable of distinguishing between the Schwarzschild and
Exponential metrics.

An important result that will be needed later follows from the
fact that the escape geodesics of photons in the exponential
metric are restricted to maximum angles of departure from the
vertical that are proportional to $e^{-2u}/r$. Since this has a
maximum for $u=1/2$, the fraction of photons emitted isotropically
from within $r < 2R_g$ that can escape is
\begin{equation}
f_{esc} = [2u(r)e^{(1-2u(r))}]^2
\end{equation}

\section{MECO in the Exponential Metric}

The general structure of the MECO is a that of a strongly
radiation pressure dominated plasma with a core temperature that
might be as large as ~ $10^{13}$K (Mitra 2006a) in GBHC. As
radiation streams outward and cools, an Eddington limit is reached
at which there is no longer sufficient outgoing radiation flux to
support baryons. This is taken to be the MECO ``surface". The
compactness there then guarantees (Cavaliere \& Morrison 1980, see
Appendix D) that photon-photon collisions would copiously produce
electron-positron pairs in the surface region. This should have
the effect of buffering the temperature at about the pair
production threshold. This makes the baryon ``surface" a phase
transition zone at the base of an electron-positron pair
atmosphere. As temperature declines further out in the pair
atmosphere the concentration of pairs decreases. Eventually a
photosphere is reached. This is taken to be a last scattering
surface, but the MECO is still so compact that only a small
fraction of isotropically emitted photons actually escape from the
photosphere.

\subsection{Central MECO Quiescent Luminosity}
Many, if not most GBHC, LMXB and AGN produce their transient high
luminosities from accretion disks. For the moment, however, the
luminosity of the MECO engine is the subject of interest.
Throughout the MECO object interior, a plasma with some baryonic
content is supported by a {\it net} outward flux of momentum via
radiation at the local Eddington limit $L_{Edd}$. At the baryon
surface this is given by
\begin{equation}
L_{Edd,s}=\frac{4 \pi G Mc (1+z_s)}{\kappa}
\end{equation}
Here $\kappa$ is the opacity of the plasma, subscript s refers to
the baryonic surface layer and $z_s$ is the gravitational redshift
at the surface given by Eq. 2. For a hydrogen plasma, $\kappa=0.4
~cm^2~ g^{-1}$ and
\begin{equation}
L_{Edd,s}=1.26\times 10^{38} m (1+z_s) ~~~~erg~ s^{-1}
\end{equation}
where $m=M/M_\odot $ is the mass in solar mass units.

Beyond the baryon surface, the pair atmosphere remains opaque. The
net outward momentum flux continues onward, but diminished by two
effects, time dilation of the rate of photon flow and
gravitational redshift of the photons. The escaping luminosity at
a location where the redshift is $z$ is thus reduced by the ratio
$(1+z)^2/(1+z_s)^2$, and the net outflow of luminosity as
radiation transits the pair atmosphere and beyond is
\begin{equation}
L_{net~out}=\frac{4 \pi G Mc (1+z)^2}{\kappa (1+z_s) }
\end{equation}
Finally, as distantly observed where $z \rightarrow 0$, the
luminosity is
\begin{equation}
L_\infty=\frac{4 \pi G M c}{\kappa (1+z_s) }
\end{equation}
For hydrogen plasma opacity of $0.4~ cm^2~ g^{-1}$, and a typical
GBHC mass of $7~M_\odot$ this equation yields $L_\infty=8.8 \times
10^{38}/(1+z_s)~ erg~ s^{-1}$. But since the quiescent luminosity
of GBHC are observed to be less than about $10^{31}~ erg~ s^{-1}$,
it is necessary to have $z_s \approx 10^8$. Even larger redshifts
are needed to satisfy the quiescent luminosity constraints for
AGN. This is extraordinary, to say the least, but no more
incredible than the $z = \infty$ of a black hole. Achieving a
redshift of $10^8$ in the exponential metric would require $u(R)
\sim 18$ and $R = R_g/18$. This would, indeed, be a very compact
object.

At the low luminosity of Eq. 8, the gravitational collapse is
characterized by an extremely long radiative lifetime, $\tau$
(RL03, Mitra 2006a) given by:
\begin{equation}
\tau= \frac{\kappa c(1+z_s)}{4 \pi G}=4.5\times 10^8 (1+z_s)~~yr
\end{equation}
With the large redshifts that would be necessary for consistency
with quiescent luminosity levels of BHC, it is clear why such a
slowly collapsing object would be called an ``eternally collapsing
object" or ECO.

At the outskirts of the pair atmosphere of an ECO the photosphere
is reached. Here the temperature and density of pairs has dropped
to a level from which photons can depart without further
scattering from positrons or electrons. Nevertheless, the redshift
is still large enough that their escape cone is small and most
isotropically emitted photons will not travel far before falling
back through the photosphere. Let the photosphere radius,
temperature and redshift be $R_p$, $T_p$ and $z_p$, respectively,
and note that the proper radiating area at the photosphere in the
isotropic metric is $4 \pi R_p^2 e^{2u(R_p)}$, $u(R_p)=R_g/R_p$
and $1+z_p = e^{u(R_p)}$. The net luminosity escaping from the
photosphere is
\begin{equation}
L_p=4(R_g/R_p)^2 e^{2(1-2*u(R_p))}4 \pi e^{2u(R_p)} R_p^2 \sigma
T_p^4=\frac{4 \pi G Mc(1+z_p)^2}{\kappa (1+z_s) }
\end{equation}
But in the pure radiation regime beyond the photosphere, the
temperature and redshift are related by
\begin{equation}
T_\infty=\frac{T}{1+z}
\end{equation}
where $T_\infty$ is the distantly observed radiation temperature.
Substituting $1+z_p = e^{u(R_p)}$ into Eq. 10 and solving for
$T_\infty$, yields
\begin{equation}
T_\infty = \frac{2.3\times 10^7}{[m(1+z_s)]^{1/4}}~~~~K
\end{equation}
Temperature given by Eq. 12 is insignificantly different (2\%
larger) than previously obtained (RL04, RL06) for a Schwarzschild
metric MECO. For example, as distantly observed, $T_\infty \sim
10^5~K$ for a $10M_\odot$ GBHC. When the radiation that escapes
the photosphere is observed at a large distance, R, where $z = 0$,
the observed fluence would be
\begin{equation}
\frac{L_\infty}{4 \pi R^2} = 4 e^2(R_g)^2 \sigma T_\infty^4
\end{equation}
The right member of Eq. 13 can be written in terms of the
distantly observed spectral distribution, for which the radiant
flux density at distance R ($ > 2R_g$) and frequency $\nu_\infty $
would be
\begin{equation}
F_{\nu_\infty} =\frac{2\pi h
\nu_\infty^3}{c^2}\frac{1}{e^{(h\nu_\infty/kT_\infty)} -
1}\frac{4e^2R_g^2}{R^2}
\end{equation}
With a factor of $4e^2$ taking the place of the number 27 found
for the Schwarzschild metric, this result is hardly changed from
what was obtained previously for a Schwarzschild metric MECO.
Previous application of Eq. 14 for the quiescent emissions of Sgr
A* showed (RL10) that the MECO model satisfied the observational
maximum luminosity constraints (Broderick, Loeb \& Narayan 2009).

\subsection{Redshift and Magnetic Field}

It seems likely that GBHC are formed via stellar core collapse.
One would expect the collapsing plasma to produce magnetic fields
via flux compression comparable to the $\sim 10^{13-14}~G$ found
in young neutron stars. The interior magnetic field near the MECO
baryon surface is assumed to be of this magnitude. The exterior
magnetic field near the base of the pair atmosphere is much larger
due to the presence of drift currents. In contrast with neutron
stars, where a relatively non-conducting crust might allow
continuity of tangential magnetic field components across the
boundary, the surface region of a MECO would be highly conductive.
Plasma in the vicinity of the surface would be subject to drift
currents proportional to ${\bf g\times B}/B^2$. It then follows
from Ampere's law that the interior and exterior fields at the
surface must differ.

Distantly observed dipole fields that would originate on a
redshifted surface have been previously examined (Baumgarte \&
Shapiro 2003, Rezzola et al. 2001). Denoting the tangential
components of magnetic field above and below the surface, S, as
$B_{\theta,S^+}$ and $B_{\theta,S^-}$, their ratio was shown to be
(RL06)
\begin{equation}
\frac{B_{\theta,S^+}}{B_{\theta,S^-}}=\frac{3(1+z_s)}{6 ln(1+z_s)}
\end{equation}
As noted above, a redshift of $\sim 10^8$ would be needed in order
to satisfy the luminosity constraints for GBHC. But if
$B_{\theta,S^-} \sim 10^{12-14}~G$, then at the surface the field
strength given by Eq. 15 could be as large as $10^{20}~G$. In
fact, it cannot be much larger than the quantum electrodynamically
determined maximum value for a NS given by $B_{\theta,S} \sim
10^{20}~G$ (Harding, A., 2003). This is because surface magnetic
fields much larger than $\sim 10^{20}~G$ would create a
spontaneous quantum electrodynamic phase transition associated
with the vacuum production of bound pairs on the MECO surface
(Zaumen 1976). For this reason, Robertson \& Leiter estimated
$B_{\theta,S^+} \sim 10^{20}~G$ and estimated $B_{\theta,S^-} \sim
10^{13.4}~G$ (Gupta, Mishra, Mishra \& Prasanna 1998). For MECO
objects differing in mass from GBHC, this last value needs to
modified by appropriate scaling. Even if the exterior field
strength is not sufficient to produce bound pairs on the MECO
surface the temperature there will be above the pair production
threshold and there will be a pair plasma in the surface region.
At the low baryon density near the surface, the interior field
should be essentially an equipartition field. Deeper within the
MECO, with densities approaching nuclear density, the field would
remain well below equipartition. At equipartition the photon
pressure in a pair plasma is proportional to $B^4$ (Pelletier \&
Marcowith 1998). It must be capable of countering the
gravitational pressure, which is proportional to the mass density.
At the compactness of a MECO, the mass density scales as $m^{-2}$.
Thus it is necessary to require $B^4 \propto m^{-2}$ or $B \propto
m^{-1/2}$. Incorporating this scaling and starting from a redshift
appropriate for a GBHC of $\sim 7M_\odot$, Eq. 15 can be solved
iteratively to yield
\begin{equation}
1+z_s = 5.7 \times 10^7 \sqrt{m}
\end{equation}
where $m=M/M_\odot$, as usual. The magnetic moment corresponding
to a $10^{20}~G$ surface magnetic field would be
\begin{equation}
\mu \sim 1.5 \times 10^{28} m^{5/2} ~~~G ~ cm^3
\end{equation}

This magnetic moment, and the spin rates given by Table 1, Eq. 6
give the MECO model a good correspondence with observations of
spectral state switches and the radio luminosities of jets for
both GBHC and AGN (RL03, RL04,RL06).

The combination of an equipartition field at the MECO surface and
a strong redshift dependence provides a measure of stability for
the MECO. If the field increased for some reason, this would cause
more pairs to be produced than just those required for the
Eddington balance of the MECO-GBHC surface region. Additional
photon pressure would then cause the MECO-GBHC surface zone to
expand. However the resultant expansion due to this process would
reduce the redshift and the magnetic field thus quenching the
vacuum production of bound pairs and allowing the MECO-GBHC
surface to contract. Similarly, if the surface zone contracted
excessively, the increase of redshift would lead to a stronger
field and more pair production, hence more radiation pressure to
support the surface zone.

Finally, note that in both GBHC and AGN, the maximum luminosity of
the low-hard x-ray states and a cutoff of radio emissions from
jets occurs at about 2\% of Eddington luminosity. To accommodate
these facts, it is necessary for the magnetic moment to scale as
$m^{5/2}$, as it does in the MECO model.

\subsection{The Photosphere}

The decline of temperature within the pair atmosphere will
eventually lead to a decrease in pair density and the occurrence
of a photosphere as a last scattering surface for outbound
photons. The escape cone at the photosphere is so small, however,
that few photons escape from there even without further
scattering. The photosphere can be found from the condition that
(Kippenhahn \& Wiggert 1990)
\begin{equation}
\int_\infty^{R_p} n_\pm \sigma_T dl = 2/3
\end{equation}
where $dl$ is an increment of proper length in the pair
atmosphere, $n_\pm$ is the combined number density of electrons
and positrons along the path, and $\sigma_T$ is the Thompson
photon-electron collision cross-section. Landau \& Lifshitz (1958)
show that
\begin{equation}
n_\pm=\frac{8\pi}{h^3}\int_{0}^{\infty}
\frac{p^2dp}{\exp{(E/kT)}+1}
\end{equation}
where p is the momentum of a particle, $E=\sqrt{p^2c^2+m_e^2c^4}$,
k is Boltzmann's constant, $h$ is Planck's constant and $m_e$, the
mass of an electron. For low temperatures such that $kT < m_ec^2$
this becomes:
\begin{equation}
n_\pm \approx 2(\frac{2 \pi m_ekT}{h^2})^{3/2} \exp{(-m_ec^2/kT)}
\end{equation}
\begin{displaymath}~~~~~~~~~~
=2.25\times 10^{30}(T_9)^{3/2} \exp{(-1/T_9)}~~cm^{-3}
\end{displaymath}
where $T_9=T/(6\times 10^9)$ K.

The path increment, $dl$, can be written in terms of changing
redshift as
\begin{equation}
dl=dr(1+z) =e^{u(r)}dr
\end{equation}
Substituting into Eq. 18 and using $r=GM/(c^2u(r))$,
$1+z=e^{u(r)}$ and $T=T_\infty (1+z)$ beyond the photosphere
provides the relation
\begin{equation}
\frac{2.22\times10^{11}m}{T_{\infty,9}}
\int_{T_{\infty,9}+\epsilon}^{T_{p9}}\frac{T_9^{3/2}e^{(-1/T_9)}dT_9}{(ln(T_9/T_{\infty,9}))^2}
= 2/3
\end{equation}
Here $\epsilon$ was taken as $10^{-6}$ just to avoid starting with
a divergent logarithm term at $T=T_\infty$. In actual fact, the
pair density is negligibly small for temperatures still well in
excess of $T=T_\infty$. Using Eqs. 11 and 16, Eq. 22 has been
numerically integrated to obtain the photosphere temperatures and
redshifts for various masses. The results are represented with
errors below 1\% for $1<m<10^{10}$ solar mass by the relations:
\begin{equation}
T_p=2.5\times 10^8 m^{-0.034}~~~~~~~~K
\end{equation}
and
\begin{equation}
1+z_p= 950 m^{0.343}
\end{equation}
These are both about half as large as values previously obtained
for a Schwarzschild metric MECO, but they likely will not provide
any basis for observational tests.

\subsection{Central MECO Luminosity Under Accretion Conditions}
It is true that a MECO will eventually achieve 100\% efficiency of
conversion of accretion mass-energy to outgoing radiation, but the
conversion takes place on the time scale of the extremely long
MECO radiative lifetime. Accreting particles that reach the
photosphere do not produce a hard radiative impact. They first
encounter soft photons, then harder photons, then
electron-positron pairs at the photosphere and they eventually
reach the baryon surface where the net outflowing luminosity is
already at the local Eddington limit rate. This provides a very
soft landing in a phase transition zone among the baryons.

The adjustment to additional mass reaching the baryon zone of the
MECO would take place on a local acoustic wave proper time scale
of order $ GM/c^3 \sim 5 m \times 10^{-6}$ s, which is about ten
orders of magnitude less than the time required for photons to
diffuse from the baryon surface through the photosphere (RL10).
Thus the MECO has ample time to adjust to any additional accreted
mass that it acquires without requiring that it be immediately
radiated away. On the other hand, the time for accreted baryons to
diffuse on into the interior is also long. At high accretion rates
they might well contribute significantly to the pressure in the
surface layers. This should be matched by an increase of radiation
pressure and additional pair generation. Because of the high
redshift, it is safe to say that none of the rapid variability of
GBHC or AGN would originate from a MECO baryon surface or
photosphere, but it might be possible that accretion would result
in an increase of luminosity flowing outward.

The only thing that is required of the MECO to maintain its
stability is for the radiation pressure in the accretion zone to
increase by enough to counter the pressure supplied by accretion.
This requirement can be examined by considering the pressure that
accreting matter might exert if stopped at the baryon surface. A
particle of proper mass $m_o$ in radial free fall would be moving
at essentially the speed of light when it reaches the surface.
According to an observer at the surface, it would have a momentum
of $\gamma m_oc$, where $\gamma=1+z_s$ is the Lorentz factor, as
can be shown from the equations of geodesic motion for an
accretion particle. If particles arrive at the locally observed
rate of $dN/dt_s$, then the quantity of momentum deposited
according to the observer at the surface would be
$(dN/dt_s)(1+z_s)m_oc$. With the substitution of $dt/(1+z_s)$ for
$dt_s$, this becomes $(dN/dt)m_o(1+z_s)^2c$. Thus the rate of
momentum transport to the surface as observed there would be
$\dot{m}_\infty (1+z_s)^2c$, where $\dot{m}_\infty = m_odN/dt$ is
the mass accretion rate as determined by a distant observer. It
would be spread over a proper area (for an exponential metric) of
$4 \pi e^{2u(R_s)}R_s^2$, which would generate accretion pressure
of $p_{accr}= \dot{m}_\infty (1+z_s)^2 c/(4 \pi e^{2u(R_s)}R_s^2)
= \dot{m}_\infty c/(4 \pi R_s^2)= \dot{m}_\infty c[ln(1+z_s)]^2/(4
\pi R_g^2)=0.11 \dot{m}_\infty [ln(1+z_s)]^2/m^2~~~erg~ cm^{-3}$.

Worst cases for this can be examined for $\dot{m}_\infty$ large
enough to make the luminosity of an accretion disk reach the level
of an Eddington limit for the MECO as judged by a distant
observer. This rate would be $\dot{m}_\infty =1.26 \times 10^{38}
m/(0.055 c^2) = 2.6\times 10^{18} m ~~~g~s^{-1}$. Highest
pressures would be generated for the lower mass MECO. Thus for
$m=10$ solar mass, the result is $p_{accr} = 10^{19} ~~~erg~
cm^{-3}$, which is a factor of $10^6$ smaller than the Eddington
limit radiation pressure of $10^{25} ~~~erg~cm^{-3}$ at the
surface temperature of $6\times 10^9~~K$. It would neither
significantly disturb the surface conditions nor be noticeable
against a much brighter external accretion flow. For the much
lower accretion rate found for Sgr A*, it was shown (RL10) that
MECO accretion luminosity was well below the observational upper
limits. Thus the claim that the low luminosity of Sgr A* is proof
of the existence of an event horizon (Broderick, Loeb \& Narayan
2009) is false.

\section{MECO - Accretion Disk Interactions}
A large number of GBHC have been found in LMXB systems. These
consist of compact objects such as neutron stars or black hole
candidates paired with dwarf stars that contribute mass that
episodically flows into the compact objects and produces x-ray
nova outbursts. These nova systems are well explained by a disk
instability model. Mass tends to accumulate in large radius
accretion disks until its own viscosity heats it to the point of
ionization. That triggers a rapid inflow that is collimated by the
gravitational field of the compact companion into an accretion
disk. The disk fills on a viscous timescale and the x-ray
luminosity goes through a series of characteristic spectral states
as the disk engages the magnetic field of the central object.

Standard gas pressure dominated `alpha' accretion disks would be
compatible with MECO. In LMXB, when the inner disk engages the
magnetosphere, the inner disk temperature is generally high enough
to produce a very diamagnetic plasma. Surface currents on the
inner disk distort the magnetopause and they also substantially
shield the outer disk such that the region of strong
disk-magnetosphere interaction is mostly confined to a ring or
torus. This shielding leaves most of the disk under the influence
of its own internal shear dynamo fields, (e.g. Balbus \& Hawley
1998, Balbus 2003). At the inner disk radius the magnetic field of
the central MECO is much stronger than the shear dynamo field
generated within the inner accretion disk.

The various spectral states begin with a true quiescence at
luminosity $L_q$, before the outburst begins. There is usually
only a factor of a few change of luminosity by the time the
accretion disk engages the magnetic field of the central object
(MECO, NS or WD) at the ``light cylinder". This marks the maximum
quiescent luminosity state at $L_{q,max}$. This is followed by a
low/hard spectral state as the outer disk encroaches into the
magnetic field. From the light cylinder radius to the ``corotation
radius", $r_c$, the x-ray luminosity of the accretion disk may
increase by a factor of $\sim 10^3 - 10^6$. At the ``corotation
radius", $r_c$, the Keplerian orbit frequency of the inner disk
matches the spin frequency of the central object and the propeller
effect ceases; at least until a significant speed differential
between the inner disk and the central object magnetic field is
re-established. When the inner disk penetrates inside corotation,
large fractions of the accreting plasma can continue on to the
central object and produce a spectral state switch to softer and
brighter emissions.

A magnetic propeller regime (Ilarianov \& Sunyaev 1975, Stella,
White \& Rosner 1986, Cui 1997, Zhang, Yu \& Zhang 1998, Campana
et al. 1998, 2002) exists until the inner disk pushes inside the
co-rotation radius, $r_c$. Plasma may depart in a jet, or as an
outflow back over the disk as it is accelerated on outwardly
curved or open magnetic field lines. Radio images of both flows
have been seen (Paragi et al. 2002). Equatorial outflows may
contribute to the low state hard spectrum by bulk Comptonization
of soft photons in the outflow, however, it seems likely that the
hard spectrum originates primarily in patchy coronal flares
(Merloni \& Fabian 2002) on a conventional geometrically thin,
optically thick accretion disk. Both outflow comptonization and
coronal flares are compatible with partial covering models for
dipping sources, in which the hard spectral region seems to be
extended (Church 2001, Church \& Balucinska-Church 2001).
Alternatively, a compact jet (Corbel \& Fender 2002) might be a
major contributor to the hard spectrum, but if so, its x-ray
luminosity must fortuitously match the power that would be
dissipated in a conventional thin disk. The power law emissions of
the low/hard state are usually cut off below $\sim 100$ keV,
consistent with a coronal temperature of $20 - 50$ keV. Bulk
comptonization would be expected to produce higher energies.

The inner disk reaching the corotation radius marks the end of the
low/hard state at its maximum luminosity, $L_c$. If there is
sufficient accretion disk pressure to push the magnetosphere
inward, intermediate or high/soft states may be produced as the
inner disk continues to fill in to the innermost marginally stable
orbit ($5.24 R_g$ for an exponential metric MECO). Inside the
marginally stable orbit plasma accelerates in a supersonic flow on
toward the central object. The brightness of the inner disk inside
the marginally stable orbit is generally diminished and may become
optically thin as the flow speed increases.

In previous work (RL02, RL03, RL04, RL06), corotation radii in the
range $30 - 70 R_g$ have been found. Since the accretion disk can
be adequately described by Newtonian gravity to radii this small,
it doesn't matter which metric we might use to describe the outer
accretion disk of a MECO. Even at the innermost marginally stable
orbit the differences between exponential and Schwarzschild
metrics are small enough to be disregarded; however, Newtonian
gravity would predict too much energy release inside the
corotation radius because it fails to account for relativistic
mass increase. With these caveats, there is no need to
recapitulate the evidence previously analyzed by Robertson \&
Leiter. The equations that they used in their analyses are
described below and recapitulated in Table 1.

\section{Characteristic Disk Luminosity Relations}

Using units of $10^{27}~G~cm^3$ for magnetic moments, $\mu$, $100$
Hz for spin, $\nu$, $10^6$ cm for radii,$r$, $10^{15}$ g/s for
accretion rates, $\dot{m}$, solar mass units, $m$, the
magnetosphere radius was found (RL04) to be:
\begin{equation}
r_m~=~8\times 10^6 {(\frac{\mu_{27}^4}{m
\dot{m}_{15}^2})}^{1/7}~~~~ cm
\end{equation}
and the disk luminosity is
\begin{equation}
L = \frac{GM\dot{m}}{2r_m}
\end{equation}
The co-rotation radius, at which disk Keplerian and magnetosphere
spins match is:
\begin{equation}
r_c = 7\times 10^6{(\frac{m}{\nu_2^2})^{1/3}} ~~~cm
\end{equation}
The low state disk luminosity at the co-rotation radius is the
maximum luminosity of the true low state and is given by:
\begin{equation}
L_c = \frac{GM\dot{m}}{2r_c}=1.5 \times 10^{34} \mu_{27}^2
{\nu_2}^3 m^{-1}~~~~erg/s
\end{equation}
The minimum high state luminosity for all accreting matter being
able to reach the central object occurs at approximately the same
accretion rate as for $L_c$ and is given by:
\begin{equation}
L_{min} = \xi \dot{m}c^2 = 1.4\times 10^{36} \xi \mu_{27}^2
\nu_2^{7/3} m^{-5/3}~~~erg/s
\end{equation}
Where $\xi \sim 0.055$ for MECO at the innermost marginally stable
orbit. The supersonic flow beyond that should generate little
luminosity and, as noted above, the central MECO would produce
very little. For neutrons stars, $\xi \sim 0.14$ for accretion
reaching the star surface.

 In true quiescence, the inner disk
radius is larger than the light cylinder radius. In NS and GBHC,
the inner disk may be ablated due to radiation from the central
object. The inner disk radius can be ablated to distances larger
than $5\times 10^4~ km$ because optically thick material can be
heated to $\sim 5000 K$ and ionized by the radiation. The maximum
disk luminosity of the true quiescent state occurs with the inner
disk radius at the light cylinder, $r_{lc}=c/\omega_s= r_m$. The
maximum luminosity of the quiescent state is typically a factor of
a few larger than the average observed quiescent luminosity.
\begin{equation}
L_{q,max} = (2.7\times 10^{30} erg/s) \mu_{27}^2 \nu_2^{9/2}
m^{1/2}
\end{equation}

The luminosity of the true quiescent state was calculated (RL03,
RL06) from the correlation of spin-down energy loss rate (Possenti
et al 2002) with x-ray luminosity in the soft x-ray band from
$\sim 0.5 - 10$ keV, assuming that the luminosity is that of an
aligned spinning magnetic dipole.
\begin{equation}
L_q = \beta \times \frac{32 \pi^4 \mu^2 \nu^4}{3c^3} = 3.8 \times
10^{33} \beta \mu_{27}^2 \nu_2^4 ~~~~~erg/s
\end{equation}
For MECO-GBHC $\beta \sim 3\times 10^{-4}$ was found (RL06).

Since the magnetic moment, $\mu_{27}$, enters each of the above
luminosity equations it can be eliminated from ratios of these
luminosities, leaving relations involving only masses and spins.
For known masses, the ratios then yield the spins. Alternatively,
if the spin is known from burst oscillations, pulses or spectral
fit determinations of $r_c$, one only needs one measured
luminosity, $L_c$ or $L_{min}$ at the end of the transition into
the soft state, to enable calculation of the remaining $\mu_{27}$
and $L_q$ (see RL02,  RL06).

For GBHC, it is a common finding that the low state inner disk
radius is much larger than that of the marginally stable accretion
disk orbit; e.g. (Markoff, Falcke \& Fender 2001, $\dot{Z}$ycki,
Done \& Smith 1997a,b, Done \& $\dot{Z}$ycki 1999, Wilson \& Done
2001). The presence of a magnetosphere is an obvious explanation.
An empty inner disk region has also been found from observations
of microlensing of the quasars Q0957+561 and Q2237+0305 (Schild,
Leiter \& Robertson 2006, 2008). These observed empty inner disk
structures are consistent with the strongly magnetic MECO model
but do not accord with standard thin disk models of accretion
flows into a black hole. Finally, note that if an inner disk
radius is found for a known mass from fitting disk spectra for
conditions near the low/high spectral state transition, the MECO
spin frequency follows from the classical Kepler relation $\nu_s=2
\pi \nu_s = \sqrt{GM/r^3}$.

To summarize, the magnetosphere/disk interaction affects nearly
all of the spectral characteristics of NS and GBHC in LMXB and
dwarf nova systems and accounts for them in a unified and complete
way, including jet formation and radio emissions. This model is
solidly consistent with accreting NS systems, for which intrinsic
magnetic moments obtained from spin-down measurements allow little
choice. Even their relatively weak magnetic fields are too strong
to ignore. Since the similar characteristics of GBHC are cleanly
explained by the same model, the MECO offers a unified theory of
LMXB phenomenology as well as extensions to AGN. Since MECO
lifetimes are orders of magnitude greater than a Hubble time, they
provide an elegant and unified framework for understanding the
broad range of observations associated with GBHC and AGN.

\section{ Low State Mass Ejection and Radio Emission}
The radio flux, $F_{\nu}$, of jet sources has a power law
dependence on frequency of the form
\begin{equation}
F_{\nu}~ \propto~ \nu^{-\alpha}
\end{equation}
It is believed to originate in jet outflows and has been shown to
be correlated with the low state x-ray luminosity [Merloni, Heinz
\& DiMatteo 2003], with $F_\nu \sim L_x^{0.7}$.  The radio
luminosity of a jet is a function of the rate at which the
magnetosphere can do work on the inner ring of the disk. This
depends on the relative speed between the magnetosphere and the
inner disk; i.e., $\dot{E} =\tau (\omega_s - \omega_k)$, or (RL04)
\begin{equation}
\dot{E} = 0.015\frac{\mu^2 \omega_s (1 -
\frac{\omega_k}{\omega_s})}{r^3}
    ~\propto~ \mu^2 M^{-3}\dot{m}_{Edd}^{6/7}\omega_s(1 - \frac{\omega_k}{\omega_s})
\end{equation}
Here $\dot{m}_{Edd}$ is the mass accretion rate divided by the
rate that
would produce luminosity at the Eddington limit for mass $M$.

Disk mass, spiraling in quasi-Keplerian orbits from negligible
speed at radial infinity must regain at least as much energy as
was radiated away in order to escape. For this to be provided by
the magnetosphere requires $\dot{E} \geq GM\dot{M}/2r$, from which
$\omega_k \leq 2\omega_s/3$. Thus the magnetosphere alone is
incapable of completely ejecting all of the accreting matter once
the inner disk reaches this limit and the radio luminosity will be
commensurately reduced and ultimately cut off at maximum x-ray
luminosity for the low state and $\omega_k=\omega_s$. For very
rapid inner disk transit through the co-rotation radius, fast
relative motion between inner disk and magnetosphere can heat the
inner disk plasma and strong bursts of radiation pressure from
either the inner disk or the central object may help to drive
large outflows while an extended jet structure is still largely
intact. This process has been calculated\footnote{though for inner
disk radii inside the marginally stable orbit [Chou \& Tajima
1999]. This would be a high state jet, but the basic mechanism
would work in low states given strong enough magnetic fields.}
using pressures and poloidal magnetic fields of unspecified
origins. A MECO is obviously capable of supplying both the field
and disk radiation pressure. The hysteresis of the low/high and
high/low state transitions may be associated with the need for the
inner disk to be completely beyond the corotation radius before a
jet can be regenerated after it has subsided.

\begin{table*}
\tiny

\begin{center}
\caption{MECO Model Equations}
\end{center}
\begin{tabular}{lll} \hline
MECO Physical Quantity & Equation & Scaling\\
\hline \\
 1.~~~Surface Redshift - (RL06 2) & $1+z_s = 5.67\times
10^7 m^{1/2}$ &
$m^{1/2}~$\\
2.~~~Quiescent Surface Luminosity $L_\infty$ - (RL06 29) &
$L_\infty=1.26\times 10^{38} m/(1+z_s)~ $~erg s$^{-1}$ &
$m^{1/2}~$
\\
3.~~~Quiescent Surface Temp $T_\infty$ - (Eq 12) &
$T_\infty=2\times 10^6/[m(1+z_s)]^{1/4}=2.3\times 10^{4}
m^{-3/8}$ K & $m^{-3/8}$\\

4.~~~Photosphere Temp. $T_p$ & $T_p=2.5 \times 10^8 m^{-0.034}$ K
&
$m^{-0.034}$\\

5.~~~Photosphere redshift $z_p$ & $1+z_p=11000 m^{0.343}$ &
$m^{0.343}$\\

6.~~~GBHC Rotation Rate, units $10^2~Hz$ - (RL06 47) &
$\nu_2=0.89[L_{q,32}/m]^{0.763}/L_{c,36}~ \approx 0.6s_1/m~Hz^a$
& $m^{-1}~$\\

7.~~~GBHC Quiescent Lum., units $10^{32}~$~erg s$^{-1}$ - (RL06
45, 46) & $L_{q,32}=1.17 [\nu_2 L_{c,36}]^{1.31} = 4.8\times
10^{-3} \mu_{27}^{2.62} \nu_2^{5.24} m^{-0.31}~ $~erg s$^{-1}$ &
$m$~~\\
8.~~~Co-rotation Radius - (RL06 40)& $R_c=7\times 10^6
[m/\nu_2^2]^{1/3}$ cm &
$m$~~\\
9.~~~Low State Luminosity at $R_c$, units $10^{36}~$~erg s$^{-1}$
(RL06 41) & $L_{c,36} = 0.015
\mu_{27,\infty}^2 \nu_2^3 / m~~$~erg s$^{-1}$ & $m$\\

10.~~Magnetic Moment, units $10^{27}~G~cm^3$ - (RL06 41, 47)&
$\mu_{27,\infty} = 8.16[L_{c,36} m/\nu_2^3]^{1/2}~~~\mu=1.7\times
10^{28}m^{5/2}~~~G~cm^3$ &
$m^{5/2}~$\\

11.~~Disk Accr. Magnetosphere Radius - (RL06 38) & $r_m(disc)=
8\times 10^6 [\mu_{27,\infty}^4/(m\dot{m}_{15}^2)]^{1/7}$ cm &
$m$~~\\

12.~~Spherical Accr. Magnetosphere Radius & $r_m(sphere)$ or axial
 $z_m(in)= 2.3\times 10^7 [\mu_{27,\infty}^4/(m\dot{m}_{15}^2)]^{1/7}$ cm &
$m$~~\\

13.~~Spher. Accr. Eq. Mag. Rad. Rotating Dipole (RTTL03) &
$r_m(out)=1.2\times 10^7[\mu_{27}^2/(\dot{m}_{15} \nu_2)]^{1/5}$ cm & $m$~~\\

14.~~Equator Poloid. Mag. Field - (RL06 41, 47,
$B_\perp=\mu_\infty/r^3$)& $B_{\infty,10} = 250m^{-3}(R_g/r)^3~ ^b
[L_{c,36}/(m^5 \nu_2^3)]^{1/2}$ gauss
& $m^{-1/2}$\\

15.~~Low State Jet Radio Luminosity - (RL04 18, 19)&
$L_{radio,36}=10^{-6.64}m^{0.84}
L_{x,36}^{2/3}[1-(L_{x,36}/L_{c,36})^{1/3}]~~ $~erg s$^{-1}$ & $m^{3/2}$~ \\
\\
\hline
\end{tabular}\\
a: $s_1$ is a small dimensionless numerical factor ($s_1 \sim 1$
for GBHC,
see text)\\
b: $R_g = GM/c^2$
\end{table*}

For the convenience of readers who might wish to relate MECO
properties to observations of compact astrophysical objects, a
number of useful relations are given in Table 1. Many of the
parameters are given in terms of quiescent x-ray luminosity $L_q$,
or the luminosity, $L_c$, at the transition high/soft
$\rightarrow$ low/hard state since these are often measurable
quantities. Mass scaling relationships for MECO are listed in the
right hand column of Table 1. They have been shown to account for
the quasar accretion structures (Schild, Leiter \& Robertson 2006,
2008) revealed by microlensing observations of the quasars
Q0957+561 and Q2237+0305. The observed structures are consistent
with the strongly magnetic MECO model but do not accord with
standard thin disk models of accretion flows into a black hole.
Since even the nearest GBHC are much too small to be resolved in
the detail shown by the microlensing techniques which were used in
the study of Q0957 and Q2237, Sgr A* is likely the only remaining
black hole candidate for which resolved images might reveal
whether or not it possesses a magnetic moment. For this reason it
is important that it be tested.

\section{Discussion and Summary}

An enormous body of physics scholarship developed primarily over
the last half century has been built on the assumption that
trapped surfaces leading to event horizons and curvature
singularities can actually exist in nature. Misner, Thorne \&
Wheeler [1973], for example in Sec. 34.6, clearly state that this
is an assumption that underlies the well-known singularity
theorems of Hawking and Penrose. Unfortunately, this assumption
has been elevated to the status of an accepted fact without proof.
Now we are finding that the idea of an event horizon presents
problems for the structure of theoretical physics (e.g.,
Abramowicz, Kluzniak \& Lasota 2014, Hawking 2014). While previous
successful ECO-MECO models have shown that giving up the idea of
event horizons does not run afoul of any astrophysical
observations, problems associated with their occurrence within
gravity theory itself may remain.

If gravity theory needs to be altered to remove the possible
occurrence of event horizons, the exponential function for
gravitational redshifts as developed above (see also Appendices A
\& B) would seem to be a good place to start. It is simple and it
rests on accepted physical principles. Regardless of anything else
that might be needed for the structure of theoretical physics,
nothing else is needed for a quantitative accounting of the
observed properties of GBHC, AGN, neutron stars (see Appendix C).
Even cosmological redshifts without ``dark energy" (Robertson
2015) can be accommodated by allowing for time dependence in the
exponential metric. As shown here, the MECO model based on the
exponential metric differs in only minor details from the previous
version that was based on a Schwarzschild metric, but it has the
virtue of resting entirely upon a metric that is incapable of
producing an event horizon.

\clearpage

\appendix
\begin{center}{\bf APPENDIX} \end{center}

\section{Einstein's elevators}

Consider two elevators, one at rest on a planet where the local
gravitational free-fall acceleration would be $g$. Let the other
be out in free space away from gravitational fields. Let it be
equipped with a rocket engine and accelerating at the same rate,
g, in the $z$ direction as determined by its on-board
accelerometer. At the time the second elevator begins to
accelerate, let a photon be emitted from a source at its floor and
let it be absorbed later in a detector in its ceiling, a distance
L away in the frame of the elevator. While the photon is in
transit, the detector acquires some speed, v, relative to inertial
frames. From the position of the detector, it is the same as if
the source were receding from it at speed v. Thus if the frequency
of the photon emitted at the floor is $\nu_0$, the detector will
detect the Doppler shifted frequency
\begin{equation}
\nu = \frac{\nu_0 (1-v/c)}{\sqrt{1-v^2/c^2}}
\end{equation} 
We can determine the speed, $v$, of the ceiling photon detector
from the special relativistic relation
\begin{equation}
a_z = \frac{dv}{dt}= \frac{a'_z}{(\gamma^3)(1+u'_z v/c^2)^3} =
\frac{g}{\gamma^3}
\end{equation} 
where $\gamma = 1/\sqrt{(1-v^2/c^2)}$ and $u'_z =0$ is the
detector speed relative to an inertial frame that is comoving and
coincident at the time the photon reaches the detector. Time
increments, $dT$ in the elevator are contracted such that $dT =
dt/\gamma $. Substituting into Eq. 35, integrating and setting T =
L/c, we obtain
\begin{equation}
v/c={\tanh(gL/c^2)}
\end{equation} 
Substituting Eq. A3 into Eq. A1, there follows
\begin{equation}
\nu = \nu_0 e^{-gL/c^2}
\end{equation} 
By the principle of equivalence an elevator at rest in an
equivalent gravitational field, would have to produce the same
frequency shift gravitationally. In this elevator, the change of
(dimensionless) gravitational potential of the ceiling relative to
the floor is, of course, $u(z=L)= gL/c^2$. So the red shift as
photons move upward from the source is given by $1+z=e^{u(z)}$. It
is necessary to require {\bf EXACT} adherence to this exponential
form because the acceleration might be large enough to make the
accelerated elevator reach a relativistic speed before the photon
arrives at the ceiling detector. This photon red shift result was
derived by Einstein in a 1907 paper (Schwartz 1977). For a time
after 1907, Einstein maintained that the metric coefficient
$g_{00}$ must be a strictly exponential function of gravitational
potential, but his final development of general relativity fails
to satisfy the requirement. The Schwarzschild metric only agrees
with the requirement to first order.

\section{Exponential Metric and General Relativity}
The (dimensionless) potential at distance $r$ from a mass, $M$, is
$u(r)=GM/c^2r$. In the exponential metric used here,
\begin{equation}
ds^2 = e^{-2u(r)}c^2 dt^2 - e^{2u(r)}(dr^2+r^2d\theta^2+r^2
sin^2(\theta)d\phi^2)
\end{equation}
the gravitational field energy density would be a tensor, $t_i^j$
with components
\begin{equation}
t_0^0=-t_1^1=t_2^2=t_3^3 = \frac{-c^4}{8\pi G} [e^{-u}\partial_r
u(r)]^2
\end{equation}
If Einstein's gravitational field equation is modified to include
this field stress-energy tensor as a source term in the right
member the field equations would become (Yilmaz 1971)
\begin{equation}
G_i^j =-(8\pi G/c^4)(T_i^j + t_i^j)
\end{equation} 
In the space exterior to $M$, where $T_i^j =0$, the $G_0^0$
equation reduces to $\nabla^2 u(r)=0$, for which $u(r)=GM/c^2r$
is, indeed, a solution and the $G_1^1,G_2^2$, and $G_3^3$
equations are satisfied.

Yilmaz theory has been developed further (e.g., Yilmaz 1975, 1977,
1980, 1992). As a theory of particles and fields it seems to be a
viable alternative to General Relativity, however, it needs
clarification of some issues concerning the field equations and
the form of $t_i^j$ in a matter continuum. For a uniform
cosmological distribution of dust particles it was recently shown
(Robertson 2015) to correctly account for the observed
cosmological redshift-luminosity relation of SNe 1a.

\section{Neutron Star Mass}
With nothing more than a plausible equation for the potential in
the interior of a mass distribution, a hydrostatic equilibrium
equation and an equation of state for neutrons, it is possible to
show that there should be a limiting mass for neutron stars that
is well below the mass of any known GBHC. The potential equation
is assumed to be
\begin{equation}
e^{-2u} \nabla^2 u = -\frac{8\pi G}{c^4} (\rho c^2 +3p)
\end{equation}
where $\rho$ is mass-energy density, $p$ is the pressure, and $3p$
its contribution to active gravitational mass density. The
hydrostatic pressure equation consistent with $g_{00}$ from the
exponential metric would be (Weinberg 1972, p127)
\begin{equation}
p' = u'(\rho c^2 + p)
\end{equation}
Where primes denote derivatives with respect to $r$. What is
needed for the solution of these equations is a neutron equation
of state and an initial trial central pressure $p(0)$. For the
latter, it was assumed that $p(0)=\rho(0)c^2/3$ corresponds to the
maximum realistic pressure because the core would be fully
relativistic under these conditions and no longer cool enough to
permit the neglect of radiation.

The AV14+UVII model equation of state of Wiringa, Fiks, and
Fabrocini was used here (Wiringa, Fiks \& Fabrocini 1988).
Densities from their Table VI were fitted to a quartic polynomial
in $p^{1/3}$. The quartic polynomial for neutron densities was of
the form $\rho=\Sigma_n a_n(p^{1/3})^n$. The coefficients are:
a$_0$= -0.9167367, a$_1$=6.960282, a$_2$=-1.462936,
a$_3$=0.267646, a$_4$=-0.0112676. Fitting errors were below one
percent over the range of densities used in these calculations.
The initial pressure and density corresponding to
$p(0)=\rho(0)c^2/3$ were $p(0)=47.7\times 10^{34}~~erg~ cm^{-3}$
and $\rho(0)= 15.9\times 10^{14}~~g~ cm^{-3}$. Using these values,
a maximum neutron star mass of $M= 1.57 M_\odot$ and surface
radius $R=re^u= 9.7~km$ was found. For a canonical $1.4 M_\odot$
star, the central pressure and density were found to be $25.9
\times 10^{34}~~erg~cm^{-3},~~12.9\times 10^{14}~g~cm^{-3}$ with a
surface radius of $12.9~km$.

After selection of an initial central density and corresponding
pressure, the solution proceeded in steps outward with
corresponding decrements of pressure until reaching $p=0$. At that
point the numerical solution was forced to match the external
potential and its derivative. Initial conditions used were
$u(0)=u'(0)=0,~~~p'(0)=0$. As a check on the validity of using the
polynomial, the field equations of the Schwarzschild metric were
solved numerically in the same stepwise fashion. Stellar radii and
masses agreed to better than one percent with those shown in Table
VI of Wiringa, Fiks \& Fabrocini. Since their values were
calculated via the Tolman, Oppenheimer, Volkoff equation which was
derived for the Schwarzschild metric, this should be expected;
however, a failure to agree would have indicated a problem with
the numerical methods used here.

\section{The $\gamma + \gamma  \leftrightarrow e^\pm$ Phase Transition}
It is well-known that a spherical volume of radius $R$
containing a luminosity $L_{\gamma}$  of  gamma ray photons with
energies $>$ 1 MeV,\footnote{1 MeV photons correspond to $T \sim
10^{10}~K$, which is only slightly beyond the pair threshold, and
easily within reach in gravitational collapse.} will become
optically thick to the $\gamma + \gamma \leftrightarrow e^\pm$
process when the optical depth is
\begin{equation}
\tau_\pm \sim n_\gamma\sigma_{\gamma\gamma}R \sim 1
\end{equation}
and
\begin{equation}
n_\gamma \sim L_\gamma/(4\pi R^2 m_ec^3)
\end{equation} 
is the number density of $\gamma$-ray photons with energies $\sim$ 1
MeV, $L_\gamma$ is the $gamma$-ray luminosity, $R$ is the radius of
the volume, $\sigma_{\gamma\gamma}$ is the pair production cross
section and $m_e$, the mass of an electron/positron.

Since $\sigma_{\gamma\gamma} \sim \sigma_T$, the Thompson
cross-section, near threshold, we can evaluate Eq. 43 as
\begin{equation}
\tau_\pm = L_\gamma \sigma_T/(4\pi R m_e c^3)
\end{equation}
Using proper length $R \sim R_g e^{u(R)}/u(R)= R_g (1+z)/ln(1+z)$,
$\sigma_T=6.65 \times 10^{-25}~~cm^2$ and $L_{30}=L_\gamma/10^{30}$,
Eq. 43 becomes
\begin{equation}
\tau_\pm= L_{30}ln(1+z)/((1+z)R_g)
\end{equation}
Using Eq. 6 for the Eddington limit luminosity of a MECO surface,
yields
\begin{equation}
\tau_\pm=1800~ ln(1+z)
\end{equation}
Thus the resultant $\gamma + \gamma \leftrightarrow e^\pm$ phase
transition in the MECO surface magnetic field, $B_S$, creates a very
optically thick pair dominated plasma.

In general, a system becomes optically thick to photon-photon pair
production when the numerical value of its compactness parameter,
$L_\gamma/R$, is $> 10^{30}~~~erg sec^{-1}~cm^{-1}$.

Taking the photon escape cone factor $\sim  1 /(1+z)^2$  into
account, the process generates a net outward non-polytropic
radiation pressure
\begin{equation}
P \propto~ ln(1 + z)m B_S^4
\end{equation}
on the MECO surface that can greatly exceed the thermal radiation
pressure at a pair production threshold temperature of $6\times 10^9
~K$. The hydrostatic equilibrium is mass scale invariantly
stabilized at the threshold of magnetically produced pairs at the
baryon surface.

\end{document}